\documentclass[fleqn,10pt]{wlscirep}
\usepackage[utf8]{inputenc}
\usepackage[T1]{fontenc}

\usepackage{subcaption}

\newcommand{\prob}[1]{\rho(#1)}
\newcommand{\condprob}[2]{\rho(#1 \left |#2 \right. )}

\newcommand{\var}{\boldsymbol{x}}
\newcommand{\stochvar}{\boldsymbol{X}}

\newcommand{\bu}{\boldsymbol{u}}
\newcommand{\buf}{\tilde{\boldsymbol{u}}}
\newcommand{\buh}{\bar{\boldsymbol{u}}}

\newcommand{\bI}{\boldsymbol{I}}
\newcommand{\bR}{\boldsymbol{R}}
\newcommand{\bW}{\boldsymbol{W}}
\newcommand{\rd}{\mathrm{d}}

\newcommand{\drift}{b}

\newcommand{\pseudotime}{\tau}

\newcommand{\SIf}{\text{SI}_{full}}
\newcommand{\SIp}{\text{SI}_{patch}}

\newcommand{\FMf}{\text{FM}_{full}}
\newcommand{\FMp}{\text{FM}_{patch}}

\newcommand{\DMf}{\text{DM}_{full}}
\newcommand{\DMp}{\text{DM}_{patch}}

\usepackage{color}
\usepackage{soul}

\title{Generative Super-Resolution of Turbulent Flows via Stochastic Interpolants}

\author[1,*]{Martin Schiødt}
\author[2]{Nikolaj T. Mücke}
\author[1]{Clara M. Velte}
\affil[1]{Technical University of Denmark, DTU Construct, Kongens Lyngby, 2800, Denmark}
\affil[2]{Centrum Wiskunde \& Informatica, Amsterdam, 1098 XG, The Netherlands}

\affil[*]{maschi@dtu.dk}

\begin{abstract}

Capturing the intricate multiscale features of turbulent flows remains a fundamental challenge due to the limited resolution of experimental data and the computational cost of high-fidelity simulations. In many practical scenarios only coarse representations of the flows are feasible, leaving crucial fine-scale dynamics unresolved. This study addresses that limitation by leveraging generative models to perform super-resolution of velocity fields and reconstruct the unresolved scales from low-resolution conditionals. In particular, the recently formalized stochastic interpolants are employed to super-resolve a case study of two-dimensional turbulence. Key to our approach is the iterative application of stochastic interpolants over local patches of the flow field, that enables efficient reconstruction without the need to process the full domain simultaneously. The patch-wise strategy is shown to yield physically consistent super-resolved flow snapshots, and key statistical quantities -- such as the kinetic energy spectrum and the spatially averaged dissipation rate -- are accurately recovered. Moreover, compared with full-field reconstruction, the patch-wise approach produces higher-quality super-resolutions, and, in general, stochastic interpolants are observed to outperform contesting generative models across a range of metrics. These results establish stochastic interpolants as a viable tool for super-resolving turbulent flows and highlight their potential for future applications.
\end{abstract}
\begin{document}

\flushbottom
\maketitle
%
%
\thispagestyle{empty}

\section{Introduction}

Super-resolution of turbulent flows is essential for bridging the gap between the limited resolution of experimental measurements or coarse simulations and the rich, multiscale dynamics inherent to turbulence. Many practical simulations -- such as Large Eddy Simulations (LES) or low-cost numerical models -- cannot afford to resolve all relevant scales due to computational constraints \cite{goc2021large}. Super-resolution techniques enable the reconstruction of fine-scale structures from coarse data, enhancing physical fidelity and enabling accurate analysis of quantities like energy spectra, vorticity, and dissipation. This is particularly valuable for data-driven modeling, control, and diagnostics of complex fluid systems \cite{brunton2020machinereview}.

In parallel with the growing influence of machine learning in imaging and language modeling, deep learning techniques have been increasingly adopted for super-resolving turbulent flows, with studies reporting significant performance gains over conventional methods \cite{nista2024influence}. Among these, deterministically trained convolutional neural networks (CNNs) are widely used due to their strong capabilities in feature extraction. Pioneering this approach in the field of turbulence, Fukami \textit{et al.} \cite{fukami2019super, fukami2021machine} applied deep CNNs to reconstruct various two-dimensional flows. While their model recovered flow statistics, such as the kinetic energy spectrum, fairly well, it exhibited non-physical artifacts and struggled to capture small-scale structures. To address these limitations, Liu \textit{et al.} \cite{liu2020deep} incorporated temporal information as a conditional input to the model. This extension yielded improved results, but the model continued to face challenges in regions dominated by viscous effects. Zhou \textit{et al.} \cite{zhou2022robust} further enhanced the model by coupling it with an approximate deconvolution method \cite{stolz1999approximate}, and extended the analysis to a case study of three-dimensional turbulence.

Although deterministic methods, such as the aforementioned, have shown promise in super-resolving turbulent flows, recent efforts have increasingly focused on the application of generative models. Generative models constitute a class of algorithms designed to approximate the probability distribution underlying a given dataset. Once this, potentially conditional, distribution is learned, the model can synthesize new realizations by sampling from the learned distribution, yielding ensembles that are statistically consistent with the original data. Due to the stochastic nature of the sampling procedure, generative models are inherently non-deterministic. Among the most widely used generative frameworks are generative adversarial networks (GANs) \cite{goodfellow2014generative}, and diffusion models (DMs) \cite{ho2020denoising}. In the context of super-resolution, generative models produce high-resolution fields conditioned on corresponding low-resolution inputs. 

Inspired by the work of Ledig \textit{et al.}\cite{ledig2017photo}, Deng \textit{et al.} \cite{deng2019super} applied GANs to super-resolve benchmark cases of two-dimensional velocity fields. Subsequently, Subramaniam \textit{et al.} \cite{subramaniam2020turbulence} extended this methodology to reconstruct both pressure and velocity fields in three-dimensional homogeneous isotropic turbulence, enhancing the resolution from $16^3$ to $64^3$. Later, Kim \textit{et al.} \cite{kim2021unsupervised} used GANs to super-resolve slices of three-dimensional turbulent flow fields. Their results demonstrated a marked improvement in statistical accuracy relative to comparable CNN architectures. More recently, DMs have been shown to outperform GANs in augmenting incomplete or corrupted measurements of two-dimensional snapshots from three-dimensional turbulent flows \cite{li2023multi}. Furthermore, an expanding body of work has successfully employed DMs to predict and super-resolve turbulent flows under a variety of configurations \cite{kohl2023benchmarking, lienen2023zero, sardar2024spectrally}. 

In this study, we employ stochastic interpolants (SIs) \cite{albergo2022building} to perform super-resolution of the velocity field in a case study of two-dimensional turbulence. Compared to DMs, SIs offer a more direct mapping between two distributions, as their inference process is initialized with an observed data point rather than Gaussian noise. Although stochastic interpolants remain a relatively recent development, especially within the context of fluid dynamics, they have been applied in a few studies to forecast and super-resolve canonical two-dimensional flows \cite{chen2024probabilistic, mucke2025physics} and to recover state variables from sparse and noisy observations \cite{chen2025flowdas}. 

We hypothesize that SIs provide improved performance over DMs due to their direct way of connecting two arbitrary distributions. This hypothesis is empirically supported through the case study presented in this work, where we train a SI to map low-resolution samples to corresponding high-resolution samples. Moreover, to extend the applicability of SIs to more complex settings -- specifically, three-dimensional turbulence -- we introduce a patch-wise strategy that iteratively super-resolves localized subdomains of the full flow field. This localized approach effectively mitigates the computational burden associated with the increased input dimensionality that arises from finer grid resolutions, a challenge that becomes particularly acute in three-dimensional applications.

The paper is structured as follows: \autoref{sec:preliminaries} presents the problem setting, provides a brief overview of the fundamentals of SIs, and details the simulation of training and test data. Our main contribution is introduced in \autoref{sec:Method}, namely the full-field and patch-wise super-resolution methods using SIs. In \autoref{sec:results_and_discussion}, we evaluate our methodology and compare it with alternative approaches, including diffusion models and flow-matching. We examine both individual super-resolution snapshots and overall statistical performance. Finally, \autoref{sec:conclusion} summarizes our findings and conclusions.

\section{Preliminaries} \label{sec:preliminaries}
This section presents the governing equations of motion for the case study considered in the current work. It provides a brief introduction to the stochastic interpolant framework, and describes the simulation methodology. Moreover, the procedure for generating the datasets used to train and evaluate the developed models is detailed.

\subsection*{Problem setting}
Super-resolution via SIs is demonstrated on a two-dimensional Kolmogorov flow case study. The flow dynamics are governed by the incompressible Navier–Stokes equations:
\begin{subequations}
\begin{align}
    \frac{\partial \boldsymbol{u}}{\partial t} + (\boldsymbol{u} \cdot \nabla ) \boldsymbol{u} &= -\nabla p + \frac{1}{\text{Re}} \nabla ^2 \boldsymbol{u} + \boldsymbol{f}, \label{eq:NSEmomentum}\\
    \nabla \cdot \boldsymbol{u} &= 0, \label{eq:NSEcontinuity}
\end{align}
\label{eq:NSEvelocity}
\end{subequations}
where 
$\boldsymbol{u}(\boldsymbol{x},t)$ is the velocity field, $p(\boldsymbol{x},t )$ the pressure and $\boldsymbol{f}(\boldsymbol{u})$ the external forcing, specified as
\begin{align}
    \boldsymbol{f}(\boldsymbol{u}) = \sum \limits_{k=4}^{6} \sin (k y) \begin{bmatrix}
        1 \\ 0
    \end{bmatrix}
    - 0.1 \boldsymbol{u}.
\end{align}
The velocity field $\bu$ obtained through direct numerical simulation (DNS) of \eqref{eq:NSEvelocity} serves as the reference target for training and evaluating the data-driven models developed in this study. These models aim to reconstruct the statistical features of $\bu$ from a filtered counterpart, $\buf$, which retains only the large-scale flow structures. Although the SI approach is presented here as a proof of concept in a two-dimensional setting, it readily generalizes to three dimensions. Indeed, this extension is a key perspective of the present study.

\subsection*{Stochastic interpolants} \label{sec:StochasticInterpolants}

Here, we briefly outline the stochastic interpolant method, as presented in \cite{chen2024probabilistic}. The method was originally presented in \cite{albergo2022building} and expanded in \cite{chen2024probabilistic,albergo2023stochastic}. 

The SI framework provides an approach for sampling from a conditional distribution by constructing a generative model that transports a point mass to a sample from a target distribution. We aim to generate samples from the conditional distribution,
\begin{align}
\condprob{\var_1}{\var_0} = \frac{\prob{\var_0, \var_1} }{\prob{\var_0}},
\end{align}
where $\prob{\var_0}$ is the marginal distribution of $\var_0$ and $\rho(\var_0, \var_1)$ represents the joint distribution of $\var_0$ and $\var_1$. $\var_0$ samples are referred to as base samples and $\var_1$ samples are referred to as target samples.

The core of the method relies on the stochastic interpolant $\bI_{\pseudotime}$, defined as
\begin{equation}
\bI_{\pseudotime} = \alpha_{\pseudotime} \var_0 + \beta_{\pseudotime} \var_1 + \sigma_{\pseudotime} \bW_{\pseudotime}, \quad \pseudotime \in [0,1], 
\end{equation}
where $\pseudotime$ is denoted \textit{pseudo-time}. $\bW_{\pseudotime}$ is a standard Wiener process independent of $(\var_0, \var_1)$, and $\alpha_{\pseudotime}$, $\beta_{\pseudotime}$, $\sigma_{\pseudotime} \in C^1([0,1])$ are pseudo-time-dependent coefficients satisfying temporal boundary conditions:
\begin{align}
\alpha_0 = \beta_1 = 1, \quad \alpha_1 = \beta_0 = \sigma_1 = 0. \label{eq:CoefRequirements}
\end{align}
These boundary conditions ensure that $\bI_0 = \var_0$ and $\bI_1 = \var_1$, creating a bridge between the point mass at $\var_0$ and the conditional distribution $\rho(\var_1|\var_0)$. The key insight is that there exists a drift term, $\drift_{\pseudotime}$, such that the conditional distribution of $\bI_{\pseudotime}$ given $\var_0$ can be generated by solving the stochastic differential equation (SDE):
\begin{equation}
\rd\stochvar_{\pseudotime} = \drift_{\pseudotime}(\stochvar_{\pseudotime}, \var_0)\rd\pseudotime + \sigma_{\pseudotime} \rd\bW_{\pseudotime}, \quad \pseudotime \in [0,1], \quad \stochvar_{\pseudotime=0} = \var_0. \label{eq:stochastic_interpolant}
\end{equation}
In particular, samples from the distribution $\condprob{\bI_1}{\var_0}$ correspond to samples from the target distribution $\condprob{\var_1}{\var_0}$ owing to the construction of the interpolant. 

It can be shown that the drift term that provides the desired property is the unique minimizer of the objective:
\begin{equation} \label{eq:stochastic_interpolant_loss}
\arg\min_{\drift_{\pseudotime}} \mathcal{L}(\drift_{\pseudotime}) = \int_0^1 \mathbb{E}\left[||\drift_{\pseudotime}(\bI_{\pseudotime}, \var_0) - \bR_{\pseudotime}||^2\right]\rd\pseudotime,
\end{equation}
with $\bR_{\pseudotime} = \dot{\alpha}_{\pseudotime} \var_0 + \dot{\beta}_{\pseudotime} \var_1 + \dot{\sigma}_{\pseudotime} \bW_{\pseudotime}$. This objective can be estimated empirically using samples from the joint distribution, making the drift learnable using standard regression techniques with neural networks. Therefore, we parameterize the drift term as a neural network, $\drift _\theta$ with weights $\theta$ and minimize an approximation of \autoref{eq:stochastic_interpolant_loss} with respect to $\theta$:
\begin{equation}
\arg\min_{\theta} L(\theta) = \frac{1}{N_{\pseudotime} N_{\text{train}}}\sum_{i=1}^{N_{\pseudotime}} \sum_{j=1}^{N_{\text{train}}} || \drift _\theta(\bI_{\pseudotime_i}^j, \var_{0}^j, \pseudotime_i) - \bR_{\pseudotime}^j||^2, 
\quad
\bI_{\pseudotime_i}^j = \alpha_{\pseudotime_i} \var_0^j + \beta_{\pseudotime_i} \var_1^j + \sigma_{\pseudotime_i} \bW_{\pseudotime_i},
\quad
\bR_{\pseudotime_i}^j = \dot{\alpha}_{\pseudotime_i} \var_0^j + \dot{\beta}_{\pseudotime_i} \var_1^j + \dot{\sigma}_{\pseudotime_i} \bW_{\pseudotime_i},
\end{equation}
where $(\var_0^j, \var_1^j) \sim \rho(\var_0, \var_1)$, $N_{\text{train}}$ is the number of training samples, and $N_{\pseudotime}$ is the number of discrete pseudo-time points. For more details on training stochastic interpolants, see \cite{chen2024probabilistic,albergo2023stochastic}.

The architecture of $\drift _\theta$, as well as the coefficients $\alpha _{\pseudotime}$, $\beta _{\pseudotime}$, $\sigma _{\pseudotime}$, which together define our stochastic interpolant will be detailed in \autoref{sec:Method}.

\subsection*{Generating training and test sets}

In this work, stochastic interpolants are trained to reconstruct simulated velocity fields, $\bu$, by super-resolving the filtered counterpart $\buf$, which represents the corresponding low-resolution field. Thus, training and test sets are produced to consist of pairs $(\var_0, \var_1) = (\buf, \bu)$. In the current study we generate 2,000 sample pairs of $\bu$ and $\buf$ for training our models. An additional 400 sample pairs are generated for evaluating model performance. We do not create a separate validation set, as we do not perform hyperparameter tuning; instead, we adopt fixed, prior-chosen hyperparameter values throughout. The target and base samples are simulated via the procedure detailed in the following subsections.

\subsubsection*{Numerical simulation}
To produce $\bu$ we first convert \eqref{eq:NSEvelocity} to its vorticity–streamfunction formulation \cite{peyret2002spectral} and solve the governing equations on a fully periodic domain 
$(x,y) \in \Omega = [0,2\pi]^2$ using the Fourier Galerkin method \cite{kopriva2009implementing}. Here, all fields are represented as truncated Fourier series, and spatial derivatives are computed exactly in spectral space due to the periodic boundary conditions.

The non-linear convective term, typically expressed in vorticity form as $\bu \cdot \nabla \omega$ is evaluated pseudospectrally. This is done by transforming the gradient of the vorticity $\nabla \omega$ and the velocity components $\bu = (u,v)$ from spectral space to physical space using an inverse Fourier transform. The non-linear product $\bu \cdot \nabla \omega$ is then computed pointwise in physical space, and the result is transformed back to spectral space using a forward Fourier transform. This approach avoids the expensive convolution sums that would arise from computing the non-linear product directly in spectral space \cite{kopriva2009implementing}. To suppress aliasing errors in the non-linear term, a dealiasing technique, namely the 2/3-rule, is employed, where the highest one-third of wavenumbers are zeroed out after transforming back to spectral space \cite{boyd2001chebyshev}. This ensures numerical stability and accuracy in the pseudospectral evaluation.

The simulation is initialized with a random seed in spectral space, where Hermitian symmetry is enforced. After evolving the simulation to a statistically steady state, the velocity field $\bu$ is sampled on a uniform 128$\times$128 grid at temporally decorrelated intervals, determined via the autocorrelation function. The Reynolds number is set to $\text{Re} = 1000$, and time integration is performed using a fourth-order Runge–Kutta scheme with a fixed timestep $\delta t = 0.025$. For further details of the simulation, we refer to our code repository which is available online.

\subsubsection*{Producing low-resolution samples}
To generate the filtered state $\buf$ from $\bu$, a series of steps are applied. First, we apply a lowpass filter to $\bu$ with a cutoff frequency of $k_{cutoff} = 8$. This operation retains only the low-frequency modes of the velocity field, effectively removing small-scale variations. The filtering is defined by the Fourier coefficients:
\begin{align}
    \hat{\buf}_{i,k\ell} 
    =
    \begin{cases}
        \hat{\bu}_{i,k\ell}, & \text{for } |k|, |\ell| \leq k_{\text{cutoff}} \\
        0, & \text{otherwise},
    \end{cases}
\end{align}
where $\kappa = \begin{bmatrix} k & \ell\end{bmatrix}^T$ is the spectral wavenumber and subscript $i = \{1,2\}$ denotes the velocity field components $u$ and $v$. Following the filtering, the field is downsampled onto a 16$\times$16  grid by retaining every 8\textsuperscript{th} grid point in both the $x-$ and $y-$coordinates, discarding the remaining points. In essence, the velocity field on the 16$\times$16 grid represents the limited resolution data available within a coarse DNS.

\section{Methodology: Stochastic interpolants for turbulence super-resolution} \label{sec:Method}

This section develops two complementary models for turbulent flow super-resolution using stochastic interpolants: $\SIf$, which processes the entire velocity field simultaneously, and $\SIp$, which employs a patch-wise strategy designed for computational scalability. Our objective is to construct a framework that allows reconstruction of high-fidelity velocity fields from coarse experimental measurements or low-resolution DNS or LES data. The SI models are trained to approximate the conditional distribution $\condprob{\bu }{\buf}$, enabling generation of statistically consistent high-fidelity samples from coarse inputs.

A key constraint of the stochastic interpolant framework is that the base and target samples ($\var_0$ and $\var_1$) must reside in the same vector space \cite{chen2024probabilistic}. To satisfy this requirement, we upsample $\buf$ prior to training. Specifically, we learn to sample from $\condprob{\bu}{\buf} = \condprob{\bu}{\text{Up}(\buf)}$. Note that the equality holds due to the deterministic nature of the chosen upsampling operator, $\text{Up}$. We employ cubic interpolation to transform the filtered and downsampled velocity field from the 16×16 grid back to the original 128×128 resolution, though alternative interpolation schemes (e.g., linear) are equally viable.

In the following subsections, we detail the implementation of $\SIf$ and $\SIp$ for sampling from $\condprob{\bu}{\buf}$.

\subsection*{Full field super-resolution}
The full field model, $\SIf$, directly super-resolves the entire velocity field in a single forward integration of the governing SDE \eqref{eq:stochastic_interpolant}. We define the stochastic interpolant with base samples $\var_0 = \buh = \text{Up}(\buf)$ and target samples $\var_1 = \bu$.

While conceptually straightforward, this approach faces computational limitations as grid resolution increases. The SDE integration required for sample generation scales poorly with domain size, which becomes particularly problematic for three-dimensional applications where memory and computational requirements are prohibitive. These scalability constraints motivate the development of the patch-wise strategy described below.

\subsection*{Patch-wise super-resolution}
The patch-wise approach, $\SIp$, addresses the computational limitations of $\SIf$ by decomposing the super-resolution task into smaller, manageable subproblems. Rather than processing the entire domain simultaneously, $\SIp$ is applied iteratively to super-resolve localized patches of the velocity field, enabling application to high-resolution three-dimensional flows where the full field method becomes computationally intractable. Thus, where $\SIf$ is applied to reconstruct $\bu$ from $\buh$, $\SIp$ is applied to reconstruct any subfield $\bu _j$ from $\buh _j$, where
\begin{align}
    \bu _j (t, x, y) = \bu (t, x, y), \quad \buh _j (t, x, y) = \buh (t,x,y), \quad \text{ for } (x,y) \in \Omega _j,
\end{align}
and the subdomain, $\Omega _j$, is defined through the partition
\begin{align}
    \Omega = \cup _j \Omega _j, \quad \Omega _i \cap \Omega_j = \varnothing \text{ for } i \neq j,
\end{align}
with $\Omega$ denoting the full spatial domain. Although applying $\SIp$ iteratively across $\Omega$ generates statistically consistent super-resolved velocity fields, initial implementations exhibited shortcomings at patch boundaries. Here it became evident, especially when considering the vorticity field computed from the generated velocity field, that two patches were generated separately. To address this issue we expand our model to consist of two separately trained patch models when seeking the generation of a full velocity field. The first patch model, which we term the \textit{free}-generator, can be applied to super-resolve the field at any arbitrary patch, $\Omega _j$, using the low-resolution neighboring patches as conditionals. The second patch model, which we term the \textit{cond}-generator (conditional generator), may be applied to super-resolve patches, where the neighboring patches have been super-resolved using the free-generator. 
\begin{figure}
    \centering
    \includegraphics[width=0.90\linewidth]{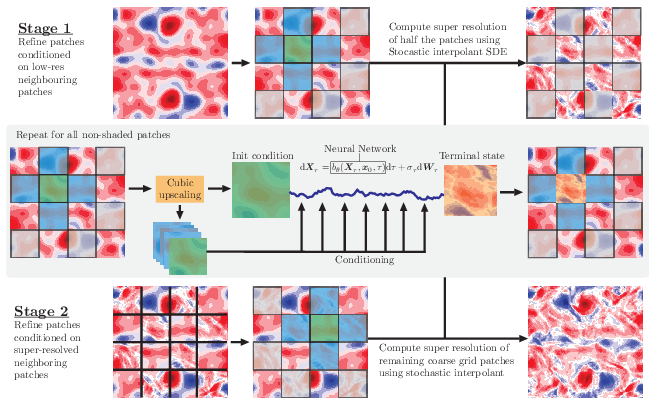}
    \caption{Visualization of the two stages in $\SIp$. Note that the states are visualized with a single channel for visual clarity. The actual data used consists of two channels -- velocity in the horizontal and vertical direction.}
    \label{fig:refining_stages}
\end{figure}

With these distinct submodules of $\SIp$, the process of super-resolving the full velocity field can be divided into two stages:

\begin{itemize}
    \item \textbf{Stage 1:} The free-generator is applied to super-resolve patches arranged in a checkerboard pattern, conditioning each patch on neighboring low-resolution patches. This yields a velocity field that is partially super-resolved (see top and middle rows of \autoref{fig:refining_stages});
    \item \textbf{Stage 2:} The cond-generator super-resolves the remaining patches conditioning on the high-resolution patches generated in Stage 1 (see middle and bottom rows of \autoref{fig:refining_stages}). Crucially, during training, this model uses neighbor patches from $\var_1$ rather than $\var_0$, emulating the process of using super-resolved data from the free-generator as boundary conditions for the cond-generator.
\end{itemize}

The two-stage approach was observed to generate more physically consistent super-resolutions than when only the free-generator was used across $\Omega$. Together, the free-generator and cond-generator thus define $\SIp$, where boundary artifacts have effectively been mitigated while maintaining the same statistical objectives as $\SIf$, but with superior computational scalability. For the current work we choose a patch size of 32$\times$32, signifying that the full velocity field may be reconstructed by super-resolving 16 separate patches.

\subsection*{Configuration of stochastic interpolants} \label{sec:Implementation}

Inspired by the results of \cite{chen2024probabilistic}, we choose, for both models, the interpolant coefficients 
\begin{align}
    \alpha _{\pseudotime} = 1 - \pseudotime, \quad \beta _{\pseudotime} = \pseudotime ^2, \quad \sigma _{\pseudotime} = 0.1(1-\pseudotime),
\end{align}
such that they satisfy \eqref{eq:CoefRequirements}. 

\subsubsection*{Network architecture}
\begin{figure}
    \centering
    \includegraphics[width=1.0\linewidth]{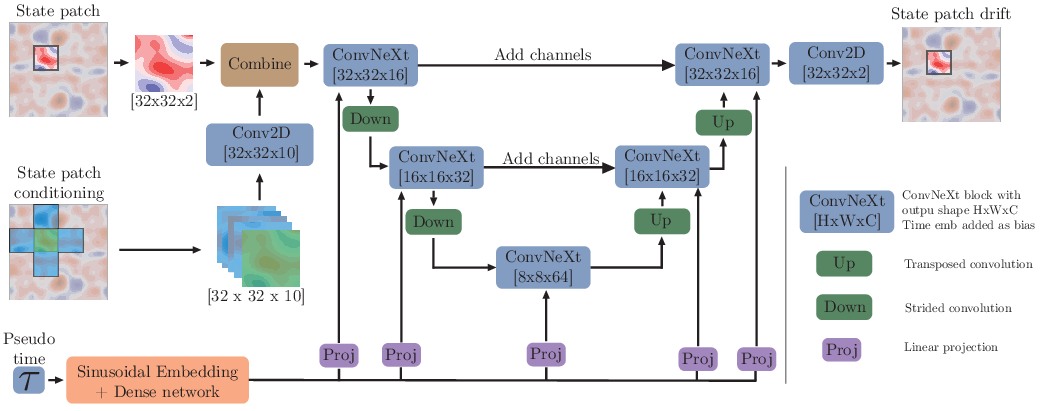}
    \caption{Visualization of the UNet architecture within the free- and cond-generators of $\SIp$.}
    \label{fig:drift_term_unet}
\end{figure}

To parameterize $\drift _\theta$ each model employs a UNet architecture, which was originally introduced by Ronneberger \textit{et al.} \cite{ronneberger2015u}. Our UNet architecture (\autoref{fig:drift_term_unet}), largely based on the approach presented in \cite{mucke2025physics}, is composed of a series of convolutional and ConvNeXt \cite{liu2022convnet} layers. The UNets of $\SIf$ and $\SIp$ differ solely in the state conditioning, where $\SIf$ takes the full $\var _0 \in \mathbb{R}^{128 \times 128 \times 2}$ as conditional input, whereas $\SIp$ takes only 5 field-patches, each of size $32 \times 32 \times 2$, as conditional input. 

Throughout the network, we employ the GELU activation function \cite{hendrycks2016gaussian}. The pseudo-time variable $\pseudotime$ is embedded using a sinusoidal positional encoding, which is then processed by a shallow neural network. This time embedding is incorporated as a conditioning input at each ConvNeXt layer as a bias within the UNet.

\subsubsection*{Divergence-free projection}
As each model is unlikely to produce a divergence free field, the output $\var _{\pseudotime = 1}$ is filtered using the Helmholtz-Hodge decomposition \cite{suda2020application}. For any field $F$ the method returns 
\begin{align}
    F _{div=0} = F - \nabla \phi, \label{eq:Helholtz1}
\end{align}
where $\phi$ solves
\begin{align}
    \nabla ^2 \phi = \nabla \cdot F. \label{eq:Helholtz2}
\end{align}
Since our velocity field is periodic, equations \eqref{eq:Helholtz1}-\eqref{eq:Helholtz2} are solved in spectral space. In other flows, the decomposition may not be as effective, and other methods to remove non-zero divergence may be needed. We refer to \cite{mucke2025physics} for a discussion of alternative projection methods.
\begin{figure}
    \centering
    \includegraphics[width=1.0\linewidth]{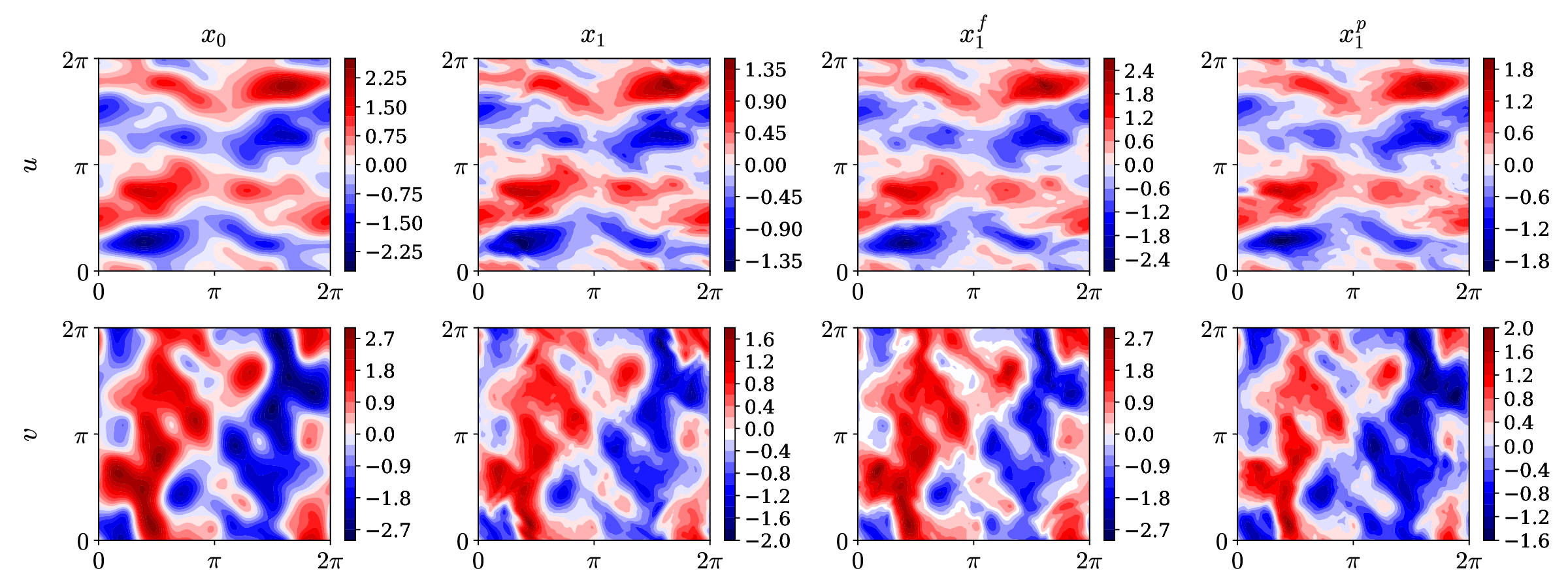}
    \caption{Velocity field for a representative snapshot. From left to right, the panels display: the low-resolution base field, the high-resolution target field, the $\SIf$ super-resolution, and the $\SIp$ super-resolution.}
    \label{fig:velocity}
\end{figure}

\subsubsection*{Training} \label{sec:training}
Each stochastic interpolant is trained over 4000 epochs using a batch size of 40 (2\% of the training set). We employ an AdamW optimizer \cite{loshchilov2017decoupled}, and apply a linear warm-up learning rate scheduler for 50 epochs. The warm-up is succeeded by a cosine annealing learning rate scheduler \cite{loshchilov2016sgdr}, with a restart period of 30 epochs. For additional implementation details, please refer to the GitHub repository linked in this work.

\section{Results \& Discussion} \label{sec:results_and_discussion}
This section presents the results of applying $\SIf$ and $\SIp$ to super-resolve the velocity field in the Kolmogorov flow case study. To represent the full field super-resolution of a snapshot we use the notation $\var _{1} ^f$ and $\var _{1} ^p$ for respectively $\SIf$ and $\SIp$. The trained models are evaluated on a test set consisting of 400 decorrelated flow snapshots. We first demonstrate that the models produce reasonable super-resolved versions of individual snapshots, followed by an analysis of statistical performance over the full test set.

\subsection*{Snapshot evaluation}
For a given snapshot, the super-resolved velocity field is inferred by forward-integrating the SDE in equation~\eqref{eq:stochastic_interpolant} using the Heun SDE integrator \cite{thygesen2023stochastic}, with 100 pseudo-timesteps and $\var_0 = \buh = \text{Up}(\buf)$ as the initial condition. The resulting high-resolution velocity field for a representative snapshot is shown in \autoref{fig:velocity}. While the models $\SIf$ and $\SIp$ are not designed to exactly reproduce $\var_1$, they are observed to successfully capture fine-scale structures that more closely resemble those seen in $\var_1$ than in the cubicly upscaled field $\var_0$.

\begin{figure}
    \centering

    \begin{subfigure}[b]{1.0\linewidth}
        \centering
        \includegraphics[width=\linewidth]{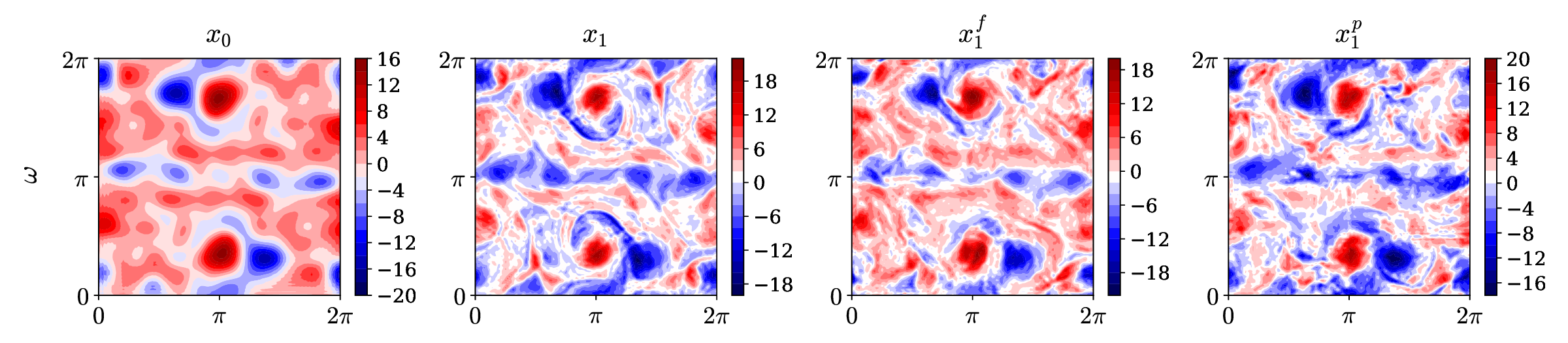}
        \caption{Vorticity.}
        \label{fig:vorticity}
    \end{subfigure}

    \begin{subfigure}[b]{1.0\linewidth}
        \centering
        \includegraphics[width=\linewidth]{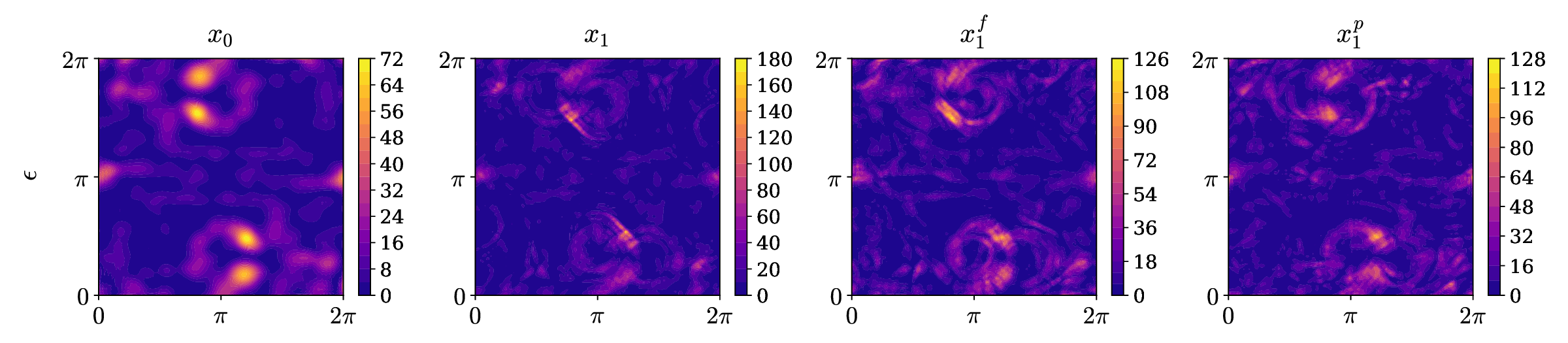}
        \caption{Dissipation rate.}
        \label{fig:dissipation}
    \end{subfigure}

    \caption{
        The (a) vorticity and (b) dissipation rate field for a representative snapshot. In each subfigure, from left to right, the panels display: the low-resolution base field, the high-resolution target field, the $\SIf$ super-resolution, and the $\SIp$ super-resolution.
    }
    \label{fig:fields_combined}
\end{figure}

The distinction becomes even more apparent when examining the vorticity field, $\omega = \nabla \times \bu$ (\autoref{fig:vorticity}). While $\var_0$ exhibits smooth, low-detail contours, both $\var_1$ and the super-resolved fields produced by the SI models display finer-scale structures. Notably, the patch-based model avoids introducing sharp discontinuities at patch boundaries (see the patch mask in \autoref{fig:refining_stages}), indicating that the cond-generator in $\SIp$ produces super-resolved patches that are consistent with those from the free-generator. Such consistency is particularly important at patch boundaries, where discontinuities in spatial gradients might otherwise arise. Fortunately, the model maintains coherent transitions across patches.

The dissipation field (\autoref{fig:dissipation}), which also depends on spatial gradients, likewise shows no discontinuities at patch boundaries. The dissipation is evaluated at each spatial point by computing
\begin{align}
    \epsilon = 2 \mu s_{ij}s_{ij}, \quad s_{ij} = \left(\frac{\partial \bu_i}{\partial \var _j} + \frac{\partial \bu _j}{\partial \var _i} \right), \quad i,j \in \{1,2\}. \label{eq:dissipation}
\end{align}
Since we solve the non-dimensionalized Navier–Stokes equations, we simply set $\mu = 1/2$ to ease the computation of $\epsilon$. As with the velocity and vorticity fields, the SI super-resolved dissipation fields more closely resembles $\var_1$ than $\var_0$.

\subsection*{Statistical performance}
We have seen that the models generate reasonable super-resolutions for a representative snapshot. We now evaluate the statistical performance over the full test set.

\begin{figure}[hbt]
    \centering
    \includegraphics[width=0.5\linewidth]{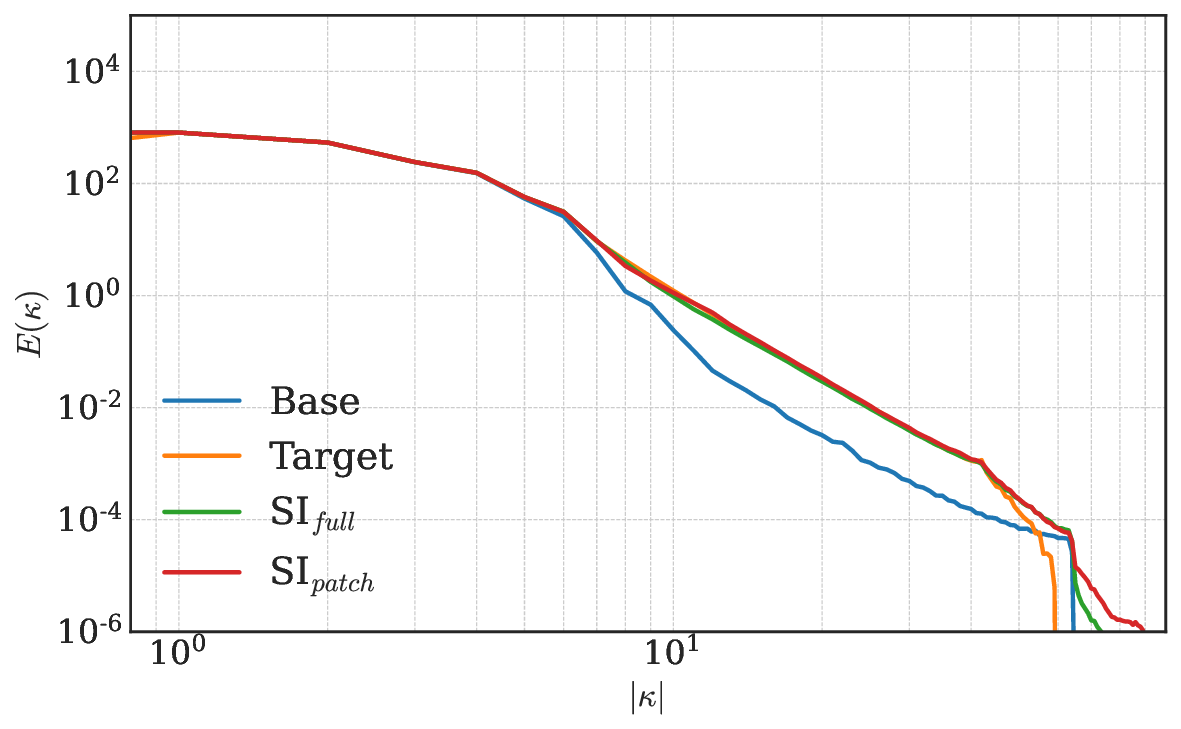}
    \caption{Radially averaged energy spectrum of the base and target fields averaged over the test set and compared to the corresponding spectra of the $\SIf$ and $\SIp$ super-resolved fields.}
    \label{fig:model_spectra}
\end{figure}

\autoref{fig:model_spectra} displays the radially averaged energy spectra of the base ($\var _0$) and target ($\var _1$) sets, and compares them to the corresponding spectra of the $\SIf$ and $\SIp$ super-resolved fields. A close alignment between the model and target spectra is observed, particularly at low to intermediate wavenumbers, highlighting a marked improvement compared to the base spectrum. At high wavenumbers the spectra diverge, however, the associated energy at these scales is minimal relative to the system energy, and the impact on overall statistical measures is therefore considered negligible. Overall, both models recover the target spectrum well. Related studies, that apply generative models for super-resolution -- such as the work by \cite{sardar2024spectrally}, which uses DMs for full-field reconstruction -- also report a divergence at high wavenumbers. 

\begin{figure}
     \centering
     \begin{subfigure}[b]{0.32\textwidth}
         \centering
         \includegraphics[width=\textwidth]{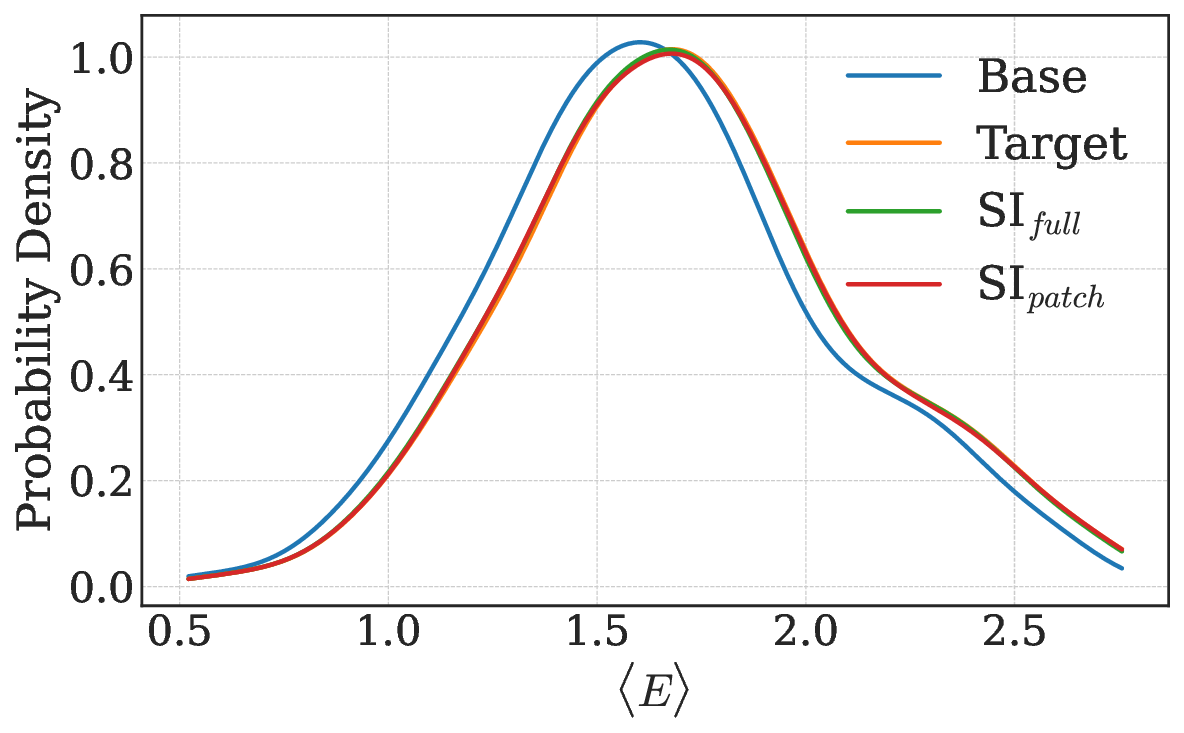}
         \caption{Kinetic energy.}
         \label{fig:KE_density}
     \end{subfigure}
     \begin{subfigure}[b]{0.32\textwidth}
         \centering
         \includegraphics[width=\textwidth]{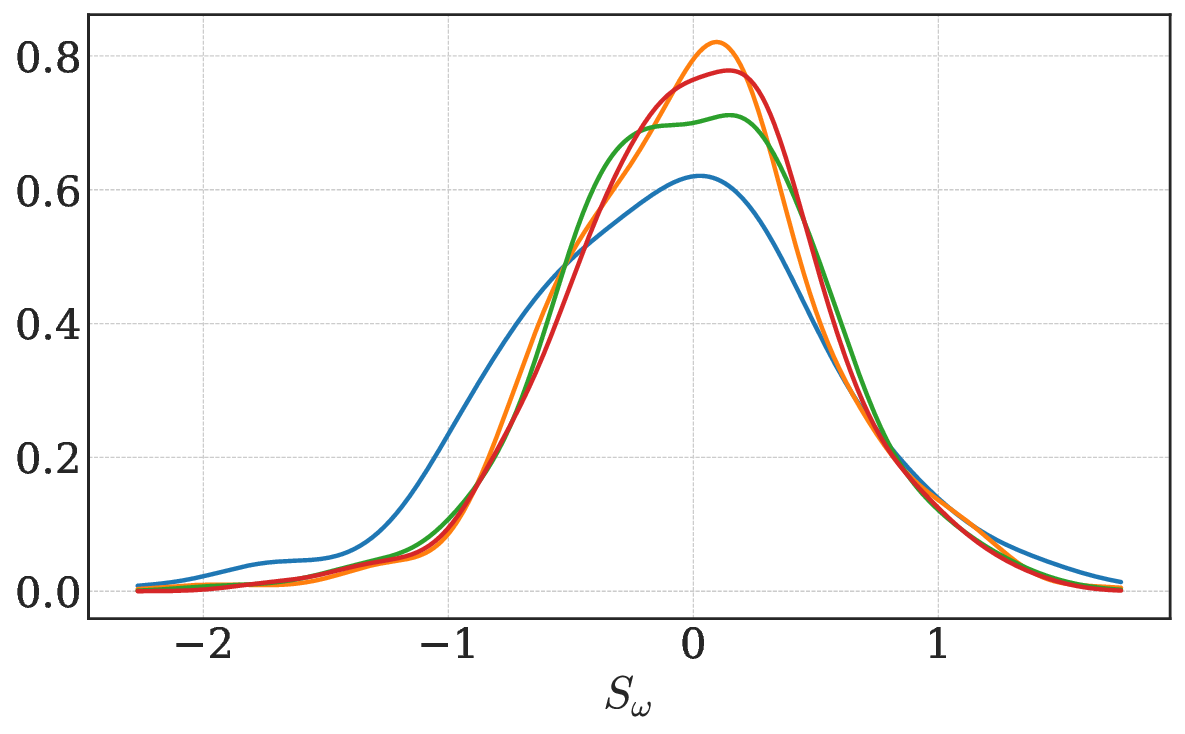}
         \caption{Vorticity skewness.}
         \label{fig:vortskew_density}
     \end{subfigure}
     \begin{subfigure}[b]{0.32\textwidth}
         \centering
         \includegraphics[width=\textwidth]{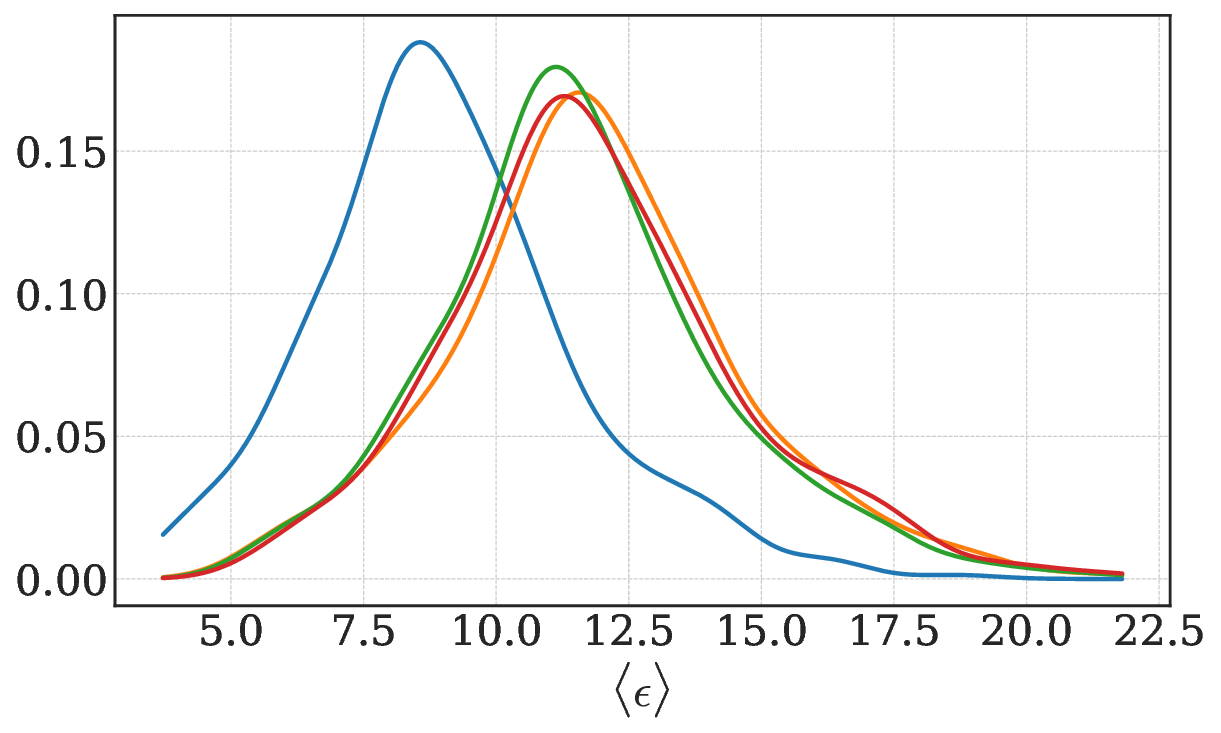}
         \caption{Dissipation rate.}
         \label{fig:dissipation_density}
     \end{subfigure}
        \caption{Probability density functions estimated using a Gaussian kernel density estimator. The plots compare the densities of $\SIf$ and $\SIp$ with those of the base, $\var _0$, and the target, $\var _1$, for the following statistics: (a) spatially averaged kinetic energy, (b) vorticity skewness, and (c) spatially averaged dissipation rate.}
        \label{fig:ProbDensity}
\end{figure}

The probability density functions shown in \autoref{fig:ProbDensity} are estimated using a Gaussian kernel density estimator \cite{scott2015multivariate, silverman2018density}. They describe the distributions of the spatially averaged kinetic energy, vorticity skewness, and spatially averaged dissipation of the base, the target, $\SIf$, and $\SIp$ fields. The average kinetic energy for each snapshot is computed as
\begin{align}
    \langle E \rangle = \left\langle u^2 + v^2 \right\rangle,
\end{align}
where $\langle \cdot \rangle$ denotes the spatial average. Equivalently, the average dissipation $\langle \epsilon \rangle$ is calculated over the dissipation field $\epsilon$ (see Equation~\eqref{eq:dissipation}). The vorticity skewness is defined as
\begin{align}
    S_\omega = \frac{\langle \omega ^3 \rangle}{\langle \omega ^2\rangle ^{3/2}}.
\end{align}
Each statistic captures a distinct aspect of the flow. As shown in \autoref{fig:KE_density}, both models accurately recover the probability density function of the kinetic energy. This outcome is expected, since the models are trained to super-resolve the velocity fields, which are directly related to the kinetic energy.
\begin{table}[ht]
    \centering
    \begin{tabular}{l|ccc|ccc}
        \toprule
        & \multicolumn{3}{c|}{KL divergence $\downarrow$} & \multicolumn{3}{c}{Wasserstein-1 distance $\downarrow$} \\
        & $\langle E \rangle$ & $S_\omega$ & $\langle \epsilon \rangle$
        & $\langle E \rangle$ & $S_\omega$ & $\langle \epsilon \rangle$ \\
        \midrule
        $\var_0$ & 0.0122 & 0.0492 & 0.5480  & 0.0594 & 0.1209 & 2.5809 \\ \hline
        $\DMf$   & 0.0001 & 0.0195 & 0.0205 & 0.0013 & 0.0862 & 0.4861  \\
        $\DMp$   & 0.0001 & 0.0099 & 0.0204 & \textbf{0.0012} & 0.0266 & 0.4908  \\ \hline 
        $\FMf$   & 0.0001 & 0.0220 & 0.0200  & 0.0025 & 0.0883 & 0.4771  \\
        $\FMp$   & 0.0001 & 0.0107 & 0.0193  & 0.0015 & 0.0243 & 0.4918  \\ \hline
        $\SIf$   & 0.0001 & 0.0068 & 0.0106  & 0.0049 & \textbf{0.0178} & 0.3013  \\
        $\SIp$   & 0.0001 & \textbf{0.0053} & \textbf{0.0040}  & 0.0031 & 0.0204 & \textbf{0.1094}  \\
        \bottomrule
    \end{tabular}
    \vspace{0.5em}
    \caption{Comparison of Kullback–Leibler divergence and Wasserstein-1 distance between the target distribution and the distributions of $\var _0$, $\DMf$, $\DMp$, $\FMf$, $\FMp$, $\SIf$ and $\SIp$, evaluated for the probability densities of spatially averaged kinetic energy, vorticity skewness, and spatially averaged dissipation. A downward-pointing arrow signifies that lower values are better. The best results are highlighted in boldface.}
    \label{tab:divergence_metrics}
\end{table}

When examining the vorticity skewness, the estimated density is not as accurately reproduced. However, the model-generated fields qualitatively approximate the behavior of the target as they produce vorticity distributions with skewness centered roughly around zero. This indicates a balanced occurrence of positive and negative vorticity, which reflects the characteristics of the target field. This behavior is also observed for the base, indicating that no apparent new information has been gained by super-resolving the low resolution field. However, when we quantify the difference between the model-generated and target probability density functions using the Kullback–Leibler (KL) divergence \cite{kullback1951information} and the Wasserstein-1 distance \cite{ramdas2017wasserstein} -- and compare it to that of the base distribution -- we find that the models approximate the vorticity skewness density roughly one order of magnitude more accurately (see \autoref{tab:divergence_metrics}). The KL divergence is defined as
\begin{align}
    D_{KL}(p || q) = \int p(x) \log \frac{p(x)}{q(x)} \mathrm{d}x,
\end{align}
where $p$ denotes the reference distribution, and $q$ is the distribution being compared or approximated. The Wasserstein-1 distance is defined as
\begin{align}
    W_1(p,q) = \inf \limits_{\pi \in \Gamma(p,q)} \mathbb{E}_{(x,y)\sim \pi} \left[ ||x-y|| \right]
\end{align}
where $\Gamma(p,q)$ is the set of all joint distributions with marginals $p$ and $q$. 

In \autoref{fig:dissipation_density}, the dissipation density functions produced by the SI models are observed to more closely approximate that of the target compared to the original low-resolution density. This is a significant result, as dissipation is a key quantity commonly used to characterize turbulent flows \cite{kolmogorov1991local}. Moreover, it is a notoriously difficult parameter to experimentally measure \cite{lai2019budgets}, and if future measurements or coarse DNS simulations can apply generative models to recover dissipation accurately, it would represent a meaningful advancement. 

\autoref{tab:divergence_metrics} also reports the KL divergence and $W_1$ distance for the densities of $\langle E \rangle$ and $\langle \epsilon \rangle$. In both cases, the SI models show an evident improvement over the base, with accuracy gains of approximately one to two orders of magnitude. Notably, $\SIp$ performs on par with, or even better than the full field model for reconstructing the considered densities. 

\begin{figure}[hbt]
    \centering
    \includegraphics[width=0.5\linewidth]{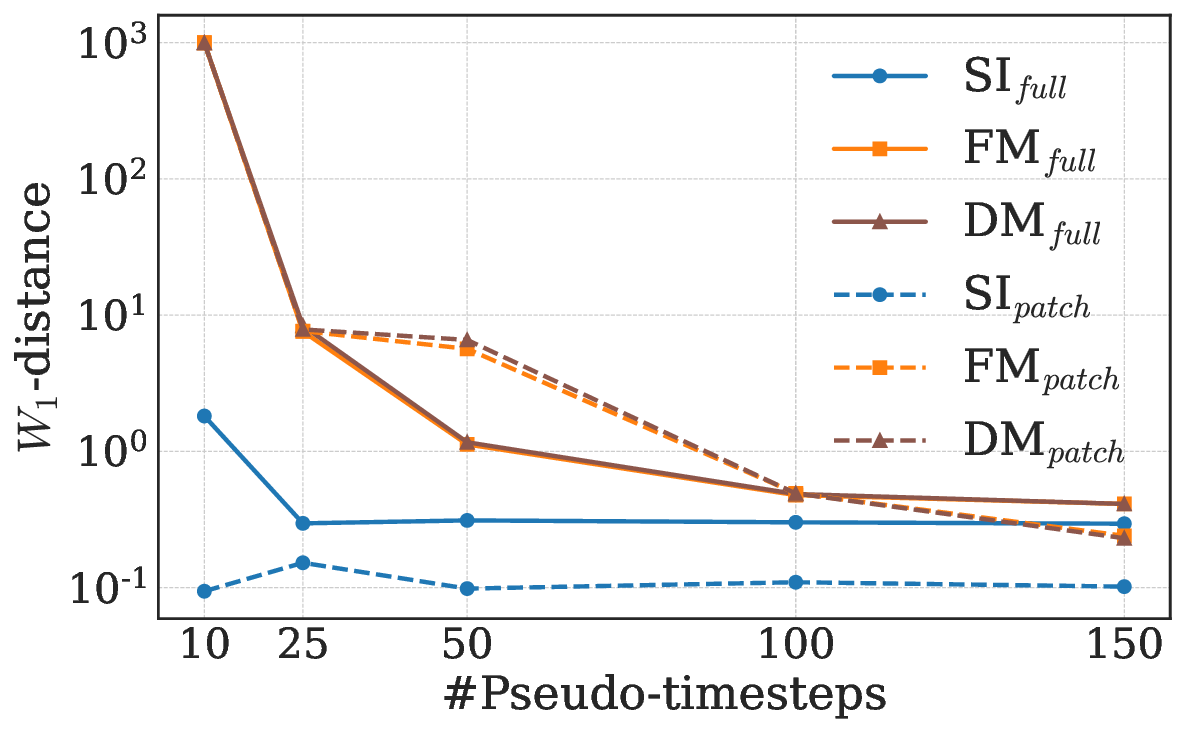}
    \caption{Convergence of the Wasserstein-1 distance of the density of $\langle \epsilon \rangle$ as a function of the number of pseudo-timesteps used to infer super-resolutions. The SI-models are observed to converge for a lower number of pseudo-timesteps compared to their contesting counterparts. For the FM- and DM-models, the $W_1$-value $10^3$ is used as a placeholder, as the models failed to produce meaningful solutions when limited to only 10 timesteps during the inference stage.}
    \label{fig:w1convergence}
\end{figure}

Finally, the table compares the performance of the SI-models to equivalent flow-matching (FM) and diffusion (DM) models (see \ref{sec:Appendix} for details). For the current configuration of using 100 pseudo-timesteps to produce super-resolutions within each model, the SI-models are observed to outperform the contesting methods in all but one of the considered metrics. This suggests that stochastic interpolants are indeed the better option for super-resolving turbulent flows. The conclusion is further supported as we consider the convergence of the $W_1$ distance for the density of $\langle \epsilon \rangle$ in \autoref{fig:w1convergence}. Here convergence is displayed as a function of the number of pseudo-timesteps used to infer super-resolutions. It is evident that the SI-models require significantly fewer timesteps to reasonably reproduce the flow statistic. Thus the SI framework is favored as the inference time decreases proportionally to the number of pseudo-timesteps.

Overall, the statistics in this section demonstrate that stochastic interpolants offer a viable approach to super-resolving turbulent flows. In particular, the results of the patch-wise strategy warrant further investigation in higher-dimensional flows.

\section{Conclusion}  \label{sec:conclusion}

We have introduced stochastic interpolants as a generative method for super-resolving fluid flows. Designed to enhance low-resolution DNS, LES, or experimental data, the approach can be applied either to reconstruct the full field in a single pass or to super-resolve smaller patches, enabling iterative recovery of the full domain or targeted regions of interest. In both configurations, the method effectively captures key flow statistics, including the energy spectrum and the probability density functions of the spatially averaged kinetic energy, vorticity skewness, and dissipation rate.

While the models developed meets the performance requirements within the studied setting, further investigation is required to evaluate the applicability to three-dimensional turbulence and generalizability to different flow regimes -- for instance, how a model trained on one type of flow behaves when applied to another. Moreover, a rigorous evaluation of inference cost relative to the computational expense of high-resolution DNS is essential to justify the use of stochastic interpolants for fluid flow super-resolution.

Compared to other state-of-the-art generative methods, such as flow-matching and diffusion models, the proposed stochastic interpolant models demonstrate superior or at least comparable performance. This highlights their potential for turbulent flow super-resolution, and offers a promising perspective for future applications.


\appendix
\renewcommand{\thesection}{Appendix \Alph{section}}
\section{Implementation of flow and diffusion model} \label{sec:Appendix}

The flow-matching (FM) and diffusion (DM) models used for comparison in this work are developed according to the framework prescribed in \cite{flowsanddiffusions2025}. Here, a flow/diffusion model is defined by the ODE/SDE used for inference (super-resolution), i.e.
\begin{align}
    \stochvar _0 \sim \mathcal{N}(0,1), \quad \mathrm{d}\stochvar_{\pseudotime} &= b_\theta(\stochvar_{\pseudotime}, \var _0, \pseudotime) \mathrm{d}t, \quad &\text{(Flow model)} \\
    \stochvar _0 \sim \mathcal{N}(0,1), \quad \mathrm{d}\stochvar_{\pseudotime} &= \tilde{b}_\theta(\stochvar_{\pseudotime}, \var _0, \pseudotime) \mathrm{d}t + \sigma _\pseudotime \mathrm{d}\bW _\pseudotime. \quad &\text{(Diffusion model)}
\end{align}
The architecture of the drift model, $b_\theta$, in $\FMf$ is identical to that used in $\SIf$ and equivalently for the patch-models. To train $b_\theta$ we follow Algorithm 3 in \cite{flowsanddiffusions2025}, where the loss 
\begin{align}
    \mathcal{L}(\theta) = \mathbb{E}_{\tau \in U_{[0,1]}, \epsilon \in \mathcal{N}(0,1)} \left[ \left|\left| b_\theta (\var _\pseudotime, \var_0, \pseudotime) - \bR_{\pseudotime} \right|\right|^2\right],
\end{align}
is minimized for batch samples of $\var _0$ and $\var _1$ under the same configurations as detailed in \autoref{sec:Method}. Here
\begin{align}
    \var _\pseudotime = \alpha _\pseudotime \epsilon + \beta _\pseudotime \var _1,
\end{align}
and for the noise-schedulers $\alpha _\pseudotime = 1-\tau^2$, $\beta _\pseudotime = \tau$, the target is given by
\begin{align}
    \bR _\pseudotime = -2\pseudotime \epsilon + \var _1.
\end{align}
After training, the flow model ODE can be solved to produce a FM super-resolution. For this purpose we use Heuns method.

For the diffusion model, we set
\begin{align}
    \tilde{b}_\theta (\stochvar _\pseudotime, \var _0, \pseudotime) = b_\theta (\stochvar _\pseudotime, \var _0, \pseudotime) + \frac{\sigma _\pseudotime}{2} s_\theta (\stochvar _\pseudotime, \var _0, \pseudotime),
\end{align}
where $b_\theta$ is the drift from the FM-models, and the score network, $s_\theta$, is evaluated directly from $b_\theta$ via
\begin{align}
    s_\theta (\stochvar _\pseudotime, \var _0, \pseudotime) = \frac{\beta _\pseudotime b_\theta (\stochvar _\pseudotime, \var _0, \pseudotime) - \dot{\beta}_\pseudotime \stochvar _\pseudotime}{\alpha _\pseudotime ^2 \dot{\beta}_\pseudotime - \beta _\pseudotime \dot{\alpha} _\pseudotime \alpha _\pseudotime}.
\end{align}
The diffusion coefficient is set to $\sigma _\pseudotime = 0.1(1-\pseudotime)$. With $\tilde{b}_\theta$ defined, the diffusion model SDE may be integrated forward in time to produce super-resolved velocity fields. For this purpose we use Heuns SDE integrator.

\section*{Acknowledgements}
C.M.V. acknowledges financial support from the European Research council: This project has received funding from the European Research Council (ERC) under the European Unions Horizon 2020 research and innovation program (grant agreement No 803419). C.M.V. and M.S. acknowledge financial support from the Poul Due Jensen Foundation: Financial support from the Poul Due Jensen Foundation (Grundfos Foundation) for this research is gratefully acknowledged. N.T.M acknowledges financial support from the National Growth Fund of the Netherlands administered by the Netherlands Organisation for Scientific Research (NWO) under the AINed XS grant NGF.1609.242.037.

\section*{Author contributions}
\textbf{M.S.}: Conceptualization, Methodology, Software, Analysis, Discussion, Writing, Visualization. \textbf{N.T.M.}: Conceptualization, Methodology, Discussion, Writing, Visualization. \textbf{C.M.V.}: Supervision, Project administration, Revisional writing.

\section*{Data availablity statement}
The code for data generation and setting up the model stochastic interpolants, is available at \url{https://github.com/martinschiodt/Turbulence_Stochastic_Interpolants}. The repository also contains scripts for training the models, performing super-resolution simulations, conducting analyses, and the implementations of the flow and diffusion models used for comparison.

\section*{Competing interests}
The authors declare no competing interests.

\end{document}